# Quantitative relationship between structural orthorhombicity, shear modulus and heat capacity anomaly of the nematic transition in iron-based superconductors

**Authors:** Joshua Javier Sanchez[1,†,*], Paul Malinowski[1], Jong-Woo Kim[2], Philip Ryan[2,3], Jiun-Haw Chu[1,*]

**Affiliation:**

[1]Department of Physics, University of Washington, Seattle, Washington 98195, USA.

[2]Advanced Photon Source, Argonne National Laboratories, Lemont, Illinois 60439, USA.

[3]School of Physical Sciences, Dublin City University, Dublin 9, Ireland.

[†]Present address: Department of Physics, Massachusetts Institute of Technology, Cambridge, Massachusetts 02139, USA.

*Correspondence to:

jhchu@uw.edu (J.-H.C)

jjsanchez2012@gmail.com (J.J.S.)

**Abstract:**

Electronic nematicity in iron pnictide materials has been extensively studied by various experimental techniques, yet its heat capacity anomaly at the phase transition has not been examined quantitatively. In this work we review the thermodynamic description of nematicity in $\text{Ba}(\text{Fe}_{1-x}\text{Co}_x)_2\text{As}_2$ using the Landau free energy, which defines the behavior of three thermodynamic quantities: the structural orthorhombicity that develops below the nematic transition, the softening shear modulus above the transition, and the discontinuous heat capacity at the transition. We derive a quantitative relationship between these three quantities, which is found to hold for a range of dopings. This result shows that the nematic transition is exceedingly well described by a mean-field model in the underdoped regime of the phase diagram.

# I. Introduction

Electronic nematicity refers to a spontaneous rotational symmetry breaking phase in solids driven by electronic correlations. In the iron pnictide superconductors, static nematic order is apparently competitive with superconductivity [1,2]; however, quantum critical nematic fluctuations likely enhance the Cooper pairing and may be the essential ingredient to the high-temperature superconductivity found in this system [3–7]. In undoped and Co-doped BaFe$_2$As$_2$, the continuous nature of the nematic transition has been verified through a variety of measured quantities, including the nematic fluctuation-driven Curie-Weiss dependence of the nematic susceptibility above the transition (from elastoresistivity, shear modulus and electronic Raman measurements [6,8–12]), the continuous 2$^{nd}$-order temperature dependence of the orthorhombicity below the transition (from x-ray diffraction and neutron scattering measurements [13–15]), and a discontinuous jump in the heat capacity at the transition (from heat capacity, elastocaloric and magnetic susceptibility measurements [16–20]). These results have all been effectively described qualitatively through a Landau mean-field model in which a phenomenological primary electronic nematic order parameter drives the transition and a linear nematic-elastic coupling induces a secondary structural orthorhombicity order parameter [10,21–23]. However, Landau theory can also be used to define specific quantitative relationships between these thermodynamic quantities.

In this work we reexamine previous experimental results and show that the quantitative relationship between different thermodynamic variables of the nematic transition can indeed be established. First, we review the Landau free energy model and show that the discontinuity in the heat capacity at the transition is directly related to the free energy coefficients that can be transformed into measurable quantities associated with the (secondary) structural orthorhombicity order parameter. We then review the structural orthorhombicity measured by x-ray diffraction, which shows a mean-field temperature dependence within the purely nematic phase. Next, above the nematic transition the shear modulus and elastoresistivity are both driven by diverging nematic fluctuations, and we demonstrate that they yield similar Curie temperatures under specific fitting assumptions. We then demonstrate the excellent agreement between the measured heat capacity discontinuity and the relation of structural quantities. Finally, the effect of disorder on the nematic transition is discussed.

## II. Landau Free Energy Overview

The Landau free energy is expressed in terms of a (primary) Ising nematic order parameter $\psi$, a (secondary) lattice orthorhombicity order parameter $\varepsilon$, and a linear nemato-elastic coupling $\lambda$ as

$$F = \frac{a\,(T-T^*)}{2}\psi^2 + \frac{b}{4}\psi^4 + \frac{C_{66,0}}{2}\varepsilon^2 - \lambda\varepsilon\psi - h\varepsilon\,. \tag{1}$$

Here, $a$ and $b$ are phenomenological parameters that control the size of the nematic order parameter and do not have any explicit microscopic definition, while $\psi$ and $\lambda$ can have a specific microscopic definition (for instance in a spin-nematic [24,25] or orbital-nematic [26,27] model) but are used in this work as model-independent thermodynamic quantities. In the absence of any coupling to the lattice ($\lambda = 0$), the nematic transition occurs at the "bare" nematic transition temperature $T^*$. As we do not use an exact microscopic definition for $\psi$, we likewise cannot put numbers to the $a$, $b$ or $\lambda$ parameters. Nonetheless, they all are well-behaved thermodynamic quantities which can be understood through the lattice-coupled quantities. The actual measurable quantities are the "bare" shear modulus $C_{66,0}$ and the nematically-coupled orthorhombicity $\varepsilon$ (in the 2-Fe unit cell this is the $B_{2g}$ orthorhombicity). We can treat ($\lambda\varepsilon$) as an effective conjugate field and tune $\psi$ through tuning $\varepsilon$, either directly with applied strain (by gluing a sample to a piezo stack or substrate) or indirectly by applied stress $h$ (with a uniaxial stress device). One further measurable quantity is the resistivity anisotropy $\eta$, with linear nemato-transport proportionality coefficient $k$:

$$\eta = \frac{\rho_{xx} - \rho_{yy}}{\rho_{xx} + \rho_{yy}} = k\psi\,. \tag{2}$$

As $\eta$ is a transport quantity and not a true thermodynamic quantity, it does not appear in the free energy. Nonetheless, as we will discuss below, it can be used to probe nematic fluctuations through the elastoresistivity technique.

We first summarize the relevant results of this model and then discuss the data interpreted through this model (see Methods of ref. [28] for further derivation notes).

Nematic fluctuations diverge from above the nematic transition temperature and are described by the "bare" nematic susceptibility $\chi$. The linear nematic-elastic coupling also results in a linear coupling of this susceptibility to the lattice, resulting in the "strain" susceptibility $\frac{d\psi}{d\varepsilon}$, which we will refer to simply as the nematic susceptibility. Both quantities diverge towards the "bare" nematic transition temperature $T^*$ with a Curie-Weiss temperature dependence:

$$\chi = \frac{1}{a(T - T^*)}, \qquad (3)$$

$$\frac{d\psi}{d\varepsilon} = \lambda \chi = \frac{\lambda}{a(T - T^*)}. \qquad (4)$$

From the fluctuation-dissipation theorem, a measurement of the susceptibility to an applied field is equivalent to a measurement of the magnitude of nematic fluctuations themselves, and so susceptibility measurements have been used widely to characterize nematic fluctuations through techniques as diverse as nuclear magnetic resonance (NMR) [29], ultrasound elastic modulus, and elastoresistivity. In all these cases, it is the lattice-coupled nematic susceptibility $\frac{d\psi}{d\varepsilon}$ that is actually measured.

A further consequence of the nematic-elastic coupling is the softening of the shear modulus $C_{66}$;

$$\frac{dh}{d\varepsilon} = C_{66} = C_{66,0} - \lambda \frac{d\psi}{d\varepsilon} = C_{66,0} - \frac{\lambda^2}{a(T - T^*)}, \qquad (5)$$

$$1 - \frac{C_{66}}{C_{66,0}} = \frac{\lambda}{C_{66,0}} \frac{d\psi}{d\varepsilon} = \frac{\lambda^2}{aC_{66,0}} \frac{1}{(T - T^*)}. \qquad (6)$$

Essentially, the nematic fluctuations soften the lattice such that a fixed stress yields an increasingly large lattice distortion on approach to the nematic transition temperature. Therefore, measurement of $C_{66}$ enables the extraction of both the Curie temperature ($T^*$) and the magnitude of the divergence ($\frac{\lambda^2}{a}$) relating to the nematic susceptibility, as well as the material-dependent and nematic-independent bare shear modulus $C_{66,0}$. The nematically-driven softening of the shear modulus towards $C_{66} = 0$ enables an infinitesimal pressure to induce a stable lattice distortion, resulting in the onset of a coupled nematic-structural phase transition. This is found to occur not at $T^*$ but at an enhanced structural transition temperature $T_S$, with the enhancement $\Delta T$ given by:

$$\Delta T = T_S - T^* = \frac{\lambda^2}{aC_{66,0}} \ . \tag{7}$$

By comparing the divergence temperature $T^*$ and the actual attained structural transition temperature $T_S$, and with knowledge of the high-temperature value of the shear modulus $C_{66,0}$, one can ascertain the ratio $\frac{\lambda^2}{a}$, which should equate the magnitude of the shear modulus divergence. This is discussed in Section 3.

Within the nematic phase ($T < T_S$) and under zero applied stress ($h = 0$), the spontaneous nematic order parameter $\psi_S$ has a mean field temperature dependence and drives a linearly proportional spontaneous structural orthorhombicity $\varepsilon_S$ as

$$\psi_S = \sqrt{\frac{a}{b}} (T_S - T)^{\frac{1}{2}}, \tag{8}$$

$$\varepsilon_S = \frac{\lambda}{C_{66,0}} \psi_S = \tilde{\varepsilon} \, (T_S - T)^{\frac{1}{2}}, \tag{9}$$

$$\tilde{\varepsilon} = \frac{\lambda}{C_{66,0}} \sqrt{\frac{a}{b}} \ . \tag{10}$$

A measurement of $\varepsilon_S$ using x-ray or neutron diffraction can only yield the magnitude $\tilde{\varepsilon}$ and the temperature dependence, which alone cannot separate out the magnitude of the nematic order parameter ($\psi = \sqrt{\frac{a}{b}}(T_S - T)^{\frac{1}{2}}$) from its proportionality to orthorhombicity $\left(\frac{\lambda}{C_{66,0}}\right)$, nor can it independently assess the enhancement of the transition temperature $\Delta T$. However, $\tilde{\varepsilon}$ provides a second measurable quantity which depends on $a$, $b$, and $\lambda$. Comparison of Eq. (10) and Eq. (7) reveals that the quantity

$$\frac{C_{66,0}\,\tilde{\varepsilon}^2}{\Delta T} = \frac{a^2}{b} \tag{11}$$

depends only on the purely nematic parameters $a$ and $b$ and not on the nematic-elastic proportionality constant $\lambda$, *even though* both $\tilde{\varepsilon}$ and $\Delta T$ would be zero in the absence of nematic-elastic coupling. To restate, the comparison of a nematically-driven structural quantity above ($C_{66}$) and below ($\varepsilon_S$) the nematic transition can be used to separate the nematic free energy parameters ($a$ and $b$) from the nematic-elastic coupling ($\lambda$).

We now show that this result further relates to the thermodynamic heat capacity. Within the nematic phase and under zero stress, we can use the mean-field nematic order parameter and the linear proportionality of nematicity to orthorhombicity ($\varepsilon_S = \frac{\lambda}{C_{66,0}}\psi_S$) to rewrite the free energy in terms of either the purely nematic or purely structural quantities as

$$F(T < T_S) = -\frac{a^2}{4b}(T_S - T)^2 = -\frac{C_{66,0}\tilde{\varepsilon}^2}{4\Delta T}(T_S - T)^2 \ . \tag{12}$$

The magnitude of the first form depends only on the purely nematic free energy parameters, $a$ and $b$, which are not directly measurable, while the magnitude of the second form depends on $C_{66,0}$, $\tilde{\varepsilon}$ and $\Delta T$ which can be extracted from measurement. We note that these magnitudes are unaffected by coupling to the lattice (the $\lambda$ term only appears in the enhanced transition temperature and not in the magnitude).

In a second order phase transition, the entropy $S$ is continuous at $T_S$ but its temperature derivative undergoes a discontinuous jump $\Delta \left(\frac{dS}{dT}\right)_{T_S}$. This is measurable as an equivalent jump in the heat capacity $\Delta C_V$, normalized by $T_S$, which yields our central theoretical result:

$$\Delta \left(\frac{dS}{dT}\right)_{T_S} = \left(\frac{d^2 F}{dT^2}\right)_{T_S^-} - \left(\frac{d^2 F}{dT^2}\right)_{T_S^+} = \frac{\Delta C_V}{T_S} = \frac{C_{66,0}\tilde{\varepsilon}^2}{2\Delta T} = \frac{a^2}{2b} \quad . \tag{13}$$

Thus, we demonstrate that in a 2$^{nd}$ order phase transition driven by a primary order parameter, the heat capacity discontinuity can be directly related to thermodynamic quantities derived from the secondary order parameter. In a system with an easily accessible primary order parameter, such as a ferromagnet, it is typically unnecessary to consider such a relation that relies on secondary order parameters. Nevertheless, this relationship allows us to access the thermodynamic behavior of the nematic order parameter without knowing the microscopic degrees of freedom by taking advantage of its coupling to the lattice. This result is generally true in any Landau model with bilinear coupled order parameters, such as in a pseudoproper ferroelastic structural transition.

In Fig. 4, we show a very good agreement between $\frac{\Delta C_V}{T_S}$ and $\frac{C_{66,0}\tilde{\varepsilon}^2}{2\Delta T}$ for the thermal fluctuation regime of the underdoped side of the $Ba(Fe_{1-x}Co_x)_2As_2$ phase diagram, confirming the validity of this result. We explore the implications of this result in the discussion section. We first review a broad set of previously reported data to extract precise values for $\frac{\Delta C_V}{T_S}$, $\tilde{\varepsilon}$, $\Delta T$ and $C_{66,0}$ across the underdoped side of the Co-doping phase diagram.

## III. X-ray diffraction of the spontaneous orthorhombicity

We first discuss evidence of a mean-field orthorhombicity within the nematic phase. Early characterizations of the structural transition using x-ray diffraction found a continuous onset of orthorhombicity at $T_S$ with either a first-order jump in orthorhombicity at $T_N$ for the parent compound and Co-dopings up to ~2.2%, or a continuous orthorhombicity with a change in slope at $T_N$ at higher dopings [1,14]. Typically, the temperature dependence of the orthorhombicity has been considered over a large temperature range across both the purely nematic phase ($T_N < T < T_S$) and the antiferromagnetic phase ($T < T_N \leq T_S$), with power law fitting used to describe the temperature evolution [1]. However, the formation of static antiferromagnetic order is expected to contribute additional effects to the temperature dependence of orthorhombic order, and within the AFM phase the orthorhombicity should not be interpreted as perfectly linear proportional to the nematicity as it can be within the paramagnetic nematic phase [11]. Therefore, we extract the mean-field magnitude of the orthorhombicity by considering only the orthorhombicity within the nematic phase, in a window of ~4K-10K below the transition.

We reexamine the x-ray diffraction data from for Co-doping values of 1.8% and 4.7% in reference [14], supplemented with our own previously unpublished data for 2.5% and 4% (see Methods). In Fig. 1(a-d) the orthorhombicity within the nematic phase is well fit to $\varepsilon = \tilde{\varepsilon}\,(T_S - T)^{\frac{1}{2}}$ in agreement with Eq. (9), despite the discontinuous jump at $T_N$ for the 1.8% Co-doped sample and the decreased slope for the 4.7% Co-doped sample at $T_N$. To assess the critical exponent of the orthorhombicity more quantitatively within the nematic phase, we plot the data on a log-log scale using the reduced temperature $\left(\frac{T_S-T}{T_S}\right)$ in Fig. 1(e). We find that for the 4 considered dopings, the nematic phase orthorhombicity can indeed be well described with a critical exponent of $\frac{1}{2}$, with $R^2 > 0.99$ for all dopings. This establishes that the orthorhombicity within the nematic phase is well described by a mean field model. We extract the orthorhombicity amplitude $\tilde{\varepsilon}$ from the fits and find that $\tilde{\varepsilon}$ smoothly decreases as the nematic transition temperature is suppressed with doping (Fig. 1(f)).

## IV. Nematic susceptibility, shear modulus, and elastoresistivity

Next, we reexamine ultrasound measurements of the shear modulus from ref. [30] and use it to extract the bare nematic transition temperature $T^*$ and to compute the nematic-elastic enhancement of the nematic/structural transition, $\Delta T = T_S - T^* = \frac{\lambda^2}{a\, C_{66,0}}$. In Fig. 2(a), the shear modulus data is plotted for three Co-dopings (0%, 3.7% and 6%) from the underdoped side of the phase diagram, along with grey lines marked the actual structural transition temperature $(T_S)$. It is seen that none of the measured dopings show a zero value at the transition itself; in fact, this is widely observed across several measurement paradigms [9,10,31–33] and has remained an open question [33]. Definitionally, the shear modulus must attain a zero value at $T_S$ for a continuous 2$^{nd}$ order structural transition to occur. Very recently [28], the elasto-XRD technique was used to demonstrate $C_{66}(T_S) = 0$ in a sample of 4% Co-doping, suggesting that in the previous measurements the non-zero shear modulus in the vicinity of the transition is likely of extrinsic origin, leading to a sub-Curie-Weiss behavior near the transition. To address this, we fit the measured shear modulus to a Curie-Weiss divergence:

$$C_{66} = C_{66,0} - \frac{A}{(T - T^*)}, \qquad (14)$$

where $A = \frac{\lambda^2}{a}$ is the Curie constant associated with the nematic susceptibility. In Fig. 2(a), this fit is applied to the three dopings under two conditions, either fitting the measured data as-is (red) or forcing the fit to pass through $C_{66} = 0$ at $T_S$ (blue). The fitted values of $A$, $C_{66,0}$ and $T^*$ are then compared with the fixed value of $T_S$ through the ratio $\frac{T_S - T^*}{A/C_{66,0}}$ which should have a value of 1 for a perfect Curie-Weiss fit. In Fig. 2(b) it is found that when forcing the fit to pass through zero, the value of $\frac{T_S - T^*}{A/C_{66,0}}$ is almost exactly 1, while for a free fit the value is progressively larger with increasing doping. This suggests that the measured values have a sub-Curie-Weiss dependence. Finally, regardless of the doping and fitting the bare shear modulus maintains a value of $C_{66,0} = 40 \pm 2\ GPa$ which we use to compare to the heat capacity in Fig. 4.

Another method to measure the nematic susceptibility is the technique of elastoresistivity [6]. Here, the $2m_{66}$ elastoresistivity coefficient is taken as directly linear to the nematic susceptibility as

$$2m_{66} = \frac{d\eta}{d\varepsilon} = k\frac{d\psi}{d\varepsilon}\ , \qquad (15)$$

where $k$ is the nemato-transport proportionality coefficient [28]. The value of $2m_{66}$ is found to diverge with an almost perfect Curie-Weiss behavior up to $T_S$ in the parent compound and with Co-doping up to approximately 2.5%, beyond which the value of $2m_{66}$ appears to dampen near $T_S$ (Fig. 2(b)). The elastoresistivity is a transport quantity, and unlike for $C_{66}$, there is not a well-defined value that $2m_{66}$ is expected to attain at $T_S$. Therefore, we cannot force the fitting to pass through a set point to address the sub-Curie-Weiss behavior and instead we fit the data up to $T_S$ (however one methodology to do systematic consistent fitting is discussed in ref. [6]).

We next compare the fitted values of $\Delta T = T_S - T^*$. For the free fitting of $C_{66}$, $\Delta T$ increases from 40 to 50 with increasing doping. However, when enforcing $C_{66}(T_S) = 0$, $\Delta T = 40$ for 0% and 3.7% Co-doping and actually decreases to 30 for 6%. Finally, the fitted $2m_{66}$ data yields values stable about $\Delta T = 25$. It is not immediately obvious why there is a discrepancy between the two measured values. However, the recent elasto-XRD paper [28] in a 4% Co-doped sample demonstrated $\Delta T = 25$ for both $C_{66}$ and $2m_{66}$, suggesting the ultrasound and other measurements undervalue the magnitude and $T^*$ of the divergence. We use both in the comparison of the heat capacity. Note that while these values are relatively stable from 0% to ~4%, the orthorhombicity magnitude decreases by about 60% from 1.8% to 4.7% Co-doping (Fig. 1(f)).

## V. Heat capacity anomaly

Finally, we reexamine the heat capacity measured for 5 samples with Co-dopings of 0%, 1.6%, 2.5%, 3.6%, and 6.1% in reference [16]. In the parent compound, any discontinuity in the heat capacity at $T_S$ is washed out by the large first-order latent heat of the antiferromagnetic transition at $T_N$, which occurs within 1K of $T_S$ (similarly, the 2$^{nd}$ order orthorhombicity that forms at $T_S$ is overwhelmed by the discontinuous jump at $T_N$, preventing a mean-field comparison of heat capacity to orthorhombicity for the parent compound). With Co-doping, the temperature splitting between $T_S$ and $T_N$ increases, allowing the heat capacity features of each transition to be cleanly separated and analyzed. Doping rapidly diminishes the jumps in heat capacity at both transitions, such that by 6.1% Co-doping the nematic transition no longer shows a definite jump. However, this makes the 6.1% Co-doped sample useful as a measure of the background phonon contribution. This background was subtracted for the 1.6%, 2.5% and 3.6% Co-doped samples to reveal the discontinuity in the heat capacity at the transition (Fig. 3). Rather than a sharp jump in $C_V$ at $T_S$ as expected theoretically, the jump is spread across a finite temperature range. To address this, we do linear fits of $C_V$ above, below and through the jump. We define $T_{S,on}$ at the onset of the jump and $T_{S,off}$ at the jump saturation. We heuristically define $T_S = \frac{1}{2}(T_{S,on} + T_{S,off})$ and the thermally normalized magnitude of the jump, $\frac{\Delta C_v}{T_S} = \frac{C_V(T_{S,off}) - C_V(T_{S,on})}{\frac{1}{2}(T_{S,on}+T_{S,off})}$. The latter is found to decrease monotonically with doping.

Fig. 4 presents our main result – the comparison between $\frac{\Delta C_v}{T_S}$ and $\frac{C_{66,0}\tilde{\varepsilon}^2}{2\,\Delta T}$. We calculate the ratio $\frac{C_{66,0}\tilde{\varepsilon}^2}{2\,\Delta T}$ using the doping-dependent values of $\tilde{\varepsilon}$ from Fig. 1(f), a doping-independent value of $C_{66,0} = 40 \pm 2$ GPa, and either $\Delta T = 25\,K$ from the $2m_{66}$ data (green) or $\Delta T = 40\,K$ from the $C_{66}$ data (blue) of Fig. 2(c). The thermally-normalized heat capacity discontinuity $\frac{\Delta C}{T_S}$ (magenta diamond) is found to be in excellent quantitative agreement with the $\Delta T = 25\ K$ values of $\frac{C_{66,0}\tilde{\varepsilon}^2}{2\ \Delta T}$ and still in good qualitative agreement with the $\Delta T = 40\ K$ values. This agreement of thermodynamic quantities indicates that the electronic nematicity in this doping range is exceedingly well described by a mean-field theory.

## VI. Discussion

Despite the success of establishing a quantitative relationship between different thermodynamic variables of the nematic phase, it is also clear that these variables appear to deviate from the perfect mean field behavior as the Cobalt concentration increases. In Fig. 5 we plot the broadening of the heat capacity jump, defined by the difference of between onset and saturate temperatures normalized by the transition temperature, $\left(\frac{T_{S,on}-T_{S,off}}{T_{S,on}+T_{S,off}}\right)$, as well as the residual value of $C_{66}$ at $T_S$. For an ideal second-order structural transition, both quantities should be zero, yet they both increase rapidly with doping, which highlights the role played by the disorder induced by Co-doping. The chemical inhomogeneity combined with the sharp change in $T_S$ with increasing doping may be a reason for the broadening of the transition. Furthermore, the Cobalt substitution may create local strain which acts as a random field to Ising nematicity and rounds off the divergence at the critical point. We note that the non-zero value of $C_{66}$ at $T_S$ leads to an apparent sub-Curie-Weiss behavior, which is strongly reminiscent of elastoresistivity $2m_{66}$ measurements in the same system. The latter has been discussed in the context of random field Ising model [6].

In conclusion, we have demonstrated that the mean-field Landau treatment of nematicity describes several thermodynamic measurement results exceedingly well. Qualitatively, it provides the correct temperature dependence of the structural orthorhombicity below the transition, the shear modulus softening above the transition, and the shape of the steplike anomaly in the heat capacity at the transition itself. Quantitatively, the magnitudes of these three quantities are shown to relate within one shared framework such that both the heat capacity and the lattice measurements give the same information about the change in entropy below the nematic transition. This formulation also yields a method to separate the nematic-elastic coupling ($\lambda$) from the phenomenological energy parameters that define the nematic order parameter size ($a$ and $b$). As the precise microscopic origin of nematicity remains under debate [34,35], it has been essential to study nematicity through its coupling to the crystal lattice. Thus, the macroscopic and thermodynamic treatment presented here serves to strengthen the foundation of this large body of work extended over numerous and diverse experimental techniques.

## VII. Methods

Single crystal samples of $Ba(Fe_{.975}Co_{.025})_2As_2$ and $Ba(Fe_{.96}Co_{.04})_2As_2$ were grown from an FeAs flux as described elsewhere [6]. X-ray diffraction measurements were performed at the Advanced Photon Source, beamline 6-ID-B, at Argonne National Laboratory. X-rays of energy 7.612 keV were used to measure the $(2\ 2\ 8)_T$ and $(0\ 0\ 8)_T$ Bragg reflections. The spontaneous orthorhombicity was determined from the peak splitting of the $(2\ 2\ 8)_T$ reflection as discussed in [14,28].

X-ray diffraction data for 1.8% and 4.7% Co-doping was obtained from ref. [14]. Ultrasound data was obtained from ref. [30]. Elastoresistivity data was obtained from ref. [6]. Heat capacity data was obtained from ref. [16].

## DATA AVAILABILITY

## ACKNOWLEDGEMENTS

We thank Rafael Fernandes, Anton Andreev, Matthias Ikeda and Elliott Rosenberg for useful conversation. This work was supported by NSF MRSEC at UW (DMR-1719797) and the Air Force Office of Scientific Research under Grant FA9550-17-1-0217 and grant FA9550-21-1-0068. J.H.C. acknowledges the support of the David and Lucile Packard Foundation, the Alfred P. Sloan foundation and the State of Washington funded Clean Energy Institute. J.L. acknowledges support from the National Science Foundation under grant no. DMR-1848269. The work performed at the Advanced Photon Source was supported by the US Department of Energy, Office of Science, and Office of Basic Energy Sciences under Contract No. DE-AC02-06CH11357. JJS acknowledges the support by the U.S. Department of Energy, Office of Science, Office of Workforce Development for Teachers and Scientists, Office of Science Graduate Student Research (SCGSR) program. The SCGSR program is administered by the Oak Ridge Institute for Science and Education (ORISE) for the DOE. ORISE is managed by ORAU under contract number DE-SC0014664. All opinions expressed in this paper are the author's and do not necessarily reflect the policies and views of DOE, ORAU, or ORISE.

## AUTHOR CONTRIBUTIONS

## ADDITIONAL INFORMATION

Competing interests: The authors declare no competing financial interests.

# Citations

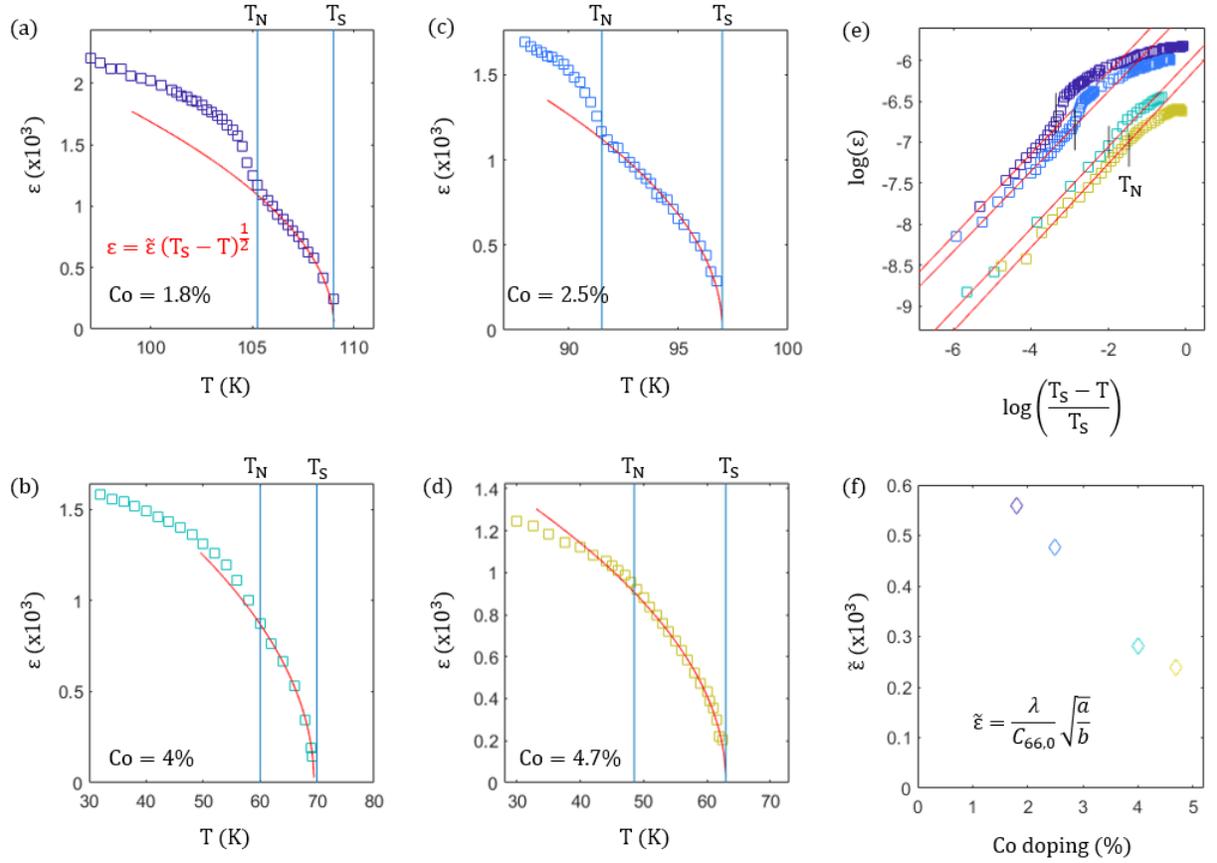

FIG. 1. (a-d) Spontaneous orthorhombicity for 4 Co-dopings. Red line is a mean-field fit $\varepsilon_S = \tilde{\varepsilon}(T_S - T)^{\frac{1}{2}}$ within the nematic phase ($T_N < T < T_S$). (e) Same dopings on log-log scale plotted vs reduced temperature. (f) Magnitude of orthorhombicity vs doping.

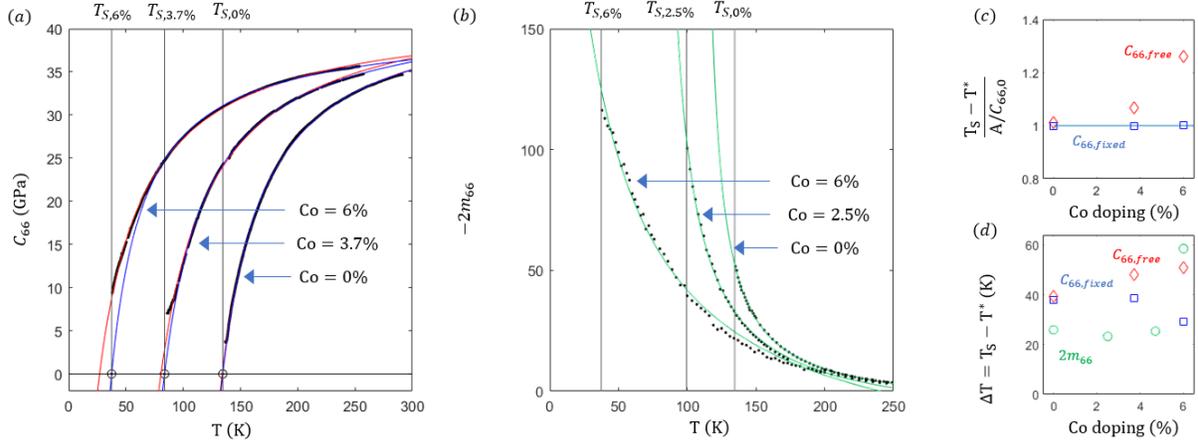

FIG. 2. (a) Shear modulus fitted to Curie-Weiss with either $C_{66}(T_S) = 0$ (blue) or nonzero (red). (b) Elastoresistivity $2m_{66}$ fit to a Curie-Weiss from $T_S$ to 250K. (c) The three fitted variables $T^*$, $A$ and $C_{66,0}$ are compared for the two types of Curie-Weiss fits in (a) and agree with the theoretical value of 1 better for the $C_{66}(T_S) = 0$ (blue) fit. (d) The extracted $\Delta T = T_S - T^*$ is ~40 for $C_{66}(T_S) = 0$ (blue square) and ~25 for $2m_{66}$ at least up to approximately 4% Co-doping (green circle).

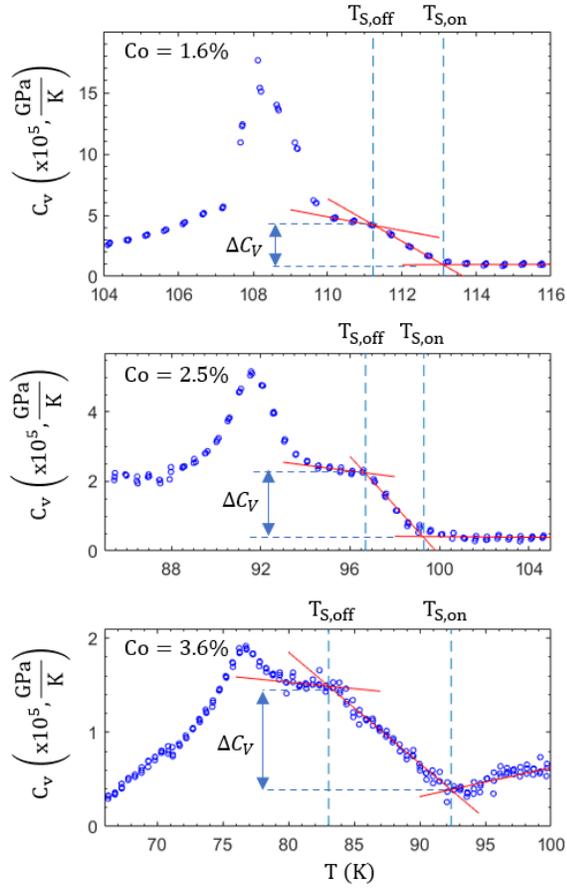

FIG. 3. Heat capacity $C_V$ with background phonon contribution subtracted for 3 doping values. The step-like increase $\Delta C_V = C_V(T_{S,off}) - C_V(T_{S,on})$ is defined from the difference between intersecting linear fits at the jump onset ($T_{S,on}$) and flattening ($T_{S,off}$) of the heat capacity. Extracted values of $\frac{\Delta C_V}{T_S}$ across the doping phase diagram presented in Fig. 4.

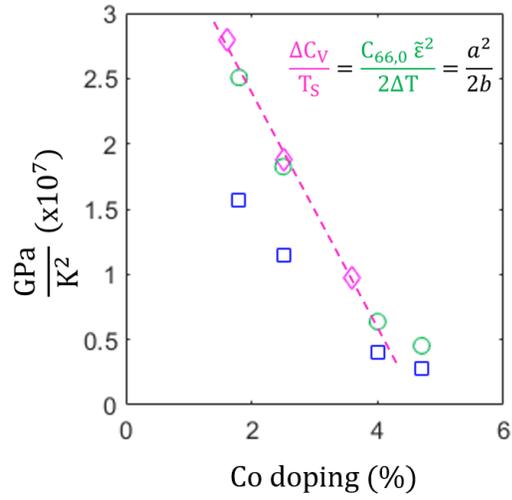

FIG. 4. Comparison of the thermally-normalized heat capacity discontinuity $\frac{\Delta C_v}{T_S}$ and the lattice quantity ratio $\frac{C_{66,0}\tilde{\varepsilon}^2}{2\,\Delta T}$ determined from either $\Delta T = 25$ (green) or $\Delta T = 40$ (blue) (see main text). Line is a guide to the eye.

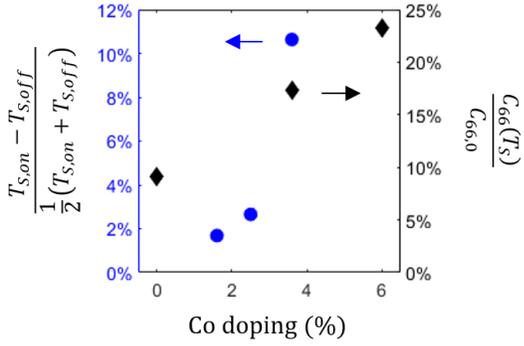

FIG. 5. Left: the finite temperature range $T_{S,on} - T_{S,off}$ of the heat capacity anomaly normalized by the transition temperature $T_S = \frac{1}{2}(T_{S,on} + T_{S,off})$ (blue circle). Right: the finite measured shear modulus at the transition temperature $C_{66}(T_S)$ normalized by the high temperature limiting value $C_{66,0}$ (black diamond).